\documentclass[prd,tightenlines]{revtex4}
\usepackage{enumerate}
\usepackage{amssymb}
\usepackage{amsmath}
\usepackage{amsfonts}
\usepackage{mathrsfs}
\usepackage{latexsym}
\usepackage{bm}
\usepackage{amscd}
\usepackage{graphicx}
\usepackage{epsfig}
\usepackage{hhline,multirow}
\usepackage{dcolumn}
\usepackage{url}
\usepackage[colorlinks=true,linkcolor=blue]{hyperref}
\usepackage{color}
\usepackage{indentfirst}
\usepackage{subfigure}
\usepackage{psfrag}
\usepackage{slashed}
\usepackage{float}
\usepackage{setspace}

\numberwithin{equation}{section}
\hypersetup{
    colorlinks=true,
    linkcolor=red,
    citecolor=blue,
}

\begin{document}

\title{On the Uniqueness of Einstein-Cartan Theory: Lagrangian, Covariant Derivative and Equation of Motion }

\author{Zi-Wei Chen}
\email{chenzw@hust.edu.cn}
\author{Run Diao}
\email{drain@hust.edu.cn}
\author{Xiang-Song Chen}
\email{cxs@hust.edu.cn}

\address{School of Physics, Huazhong University of Science and Technology, Wuhan 430074, China}

\begin{abstract}
  In the standard Einstein-Cartan theory(EC), matter fields couple to gravitation field through the Minimal Coupling Procedure(MCP), yet leaving the theory an ambiguity: applying MCP to the action or to the equation of motion would lead to different gravitational couplings. We propose a new covariant derivative to remove the ambiguity, then discuss the relation between our proposal and previous treatments on this subject.

\end{abstract}

%%%\pacs{12.20.-m; 13.40.Gp; 14.60.-z} ???????

\maketitle

\section{Introduction}
  Recall that in Einstein's General Relativity(GR) the effect of gravity is assessed on a physical system through the MCP, viz: first write down the Lagrangian or the equation of motion which holds in Special Relativity(SR), then replace the Minkowski metric $\eta_{\mu\nu}$ with general metric $g_{\mu\nu}$ and replace all the ordinary derivatives $\partial$ with covariant derivatives $\widetilde{\nabla}$\cite{Weinberg:1972kfs}. Let us  take the free scalar field Lagrangian   $\mathcal{I}^\phi_{SR}=\int d^{4}x\sqrt{g}[-\frac{1}{2}\partial_\mu\phi\partial^\mu\phi-\frac{1}{2}m^2\phi^2]$ for example. The Lagrangian in Riemann Spacetime is obtained by MCP and subsequently the equation of motion by variation over $\phi$:
\begin{equation} \label{1} 
\left\{
\begin{aligned}
&\mathcal{I}^\phi_{GR}=\int d^{4}x\sqrt{g}[-\frac{1}{2}\widetilde{\nabla}_\mu\phi\widetilde{\nabla}^\mu\phi-\frac{1}{2}m^2\phi^2] \\
&\widetilde{\nabla}_\mu\widetilde{\nabla}^\mu\phi-m^2\phi=0.
\end{aligned}
\right.
\end{equation}
Note that in (\ref{1}) the equation of motion is derived through the MCP action and coincide with the one that derived through MCP from the special-relativistic equation of motion. This is natural because the theory should not depend on whether we regard the Lagrangian or the equation of motion as a starting point of MCP. However the case becomes subtle in EC\cite{Hehl:1976kj}. When working in Riemann-Cartan Spacetime, the MCP requires replacing ordinary derivatives with covariant derivatives $\nabla$ that contain torsion. Following the similar steps in GR we get:
\begin{equation} \label{b20} 
\left\{
\begin{aligned}
&\mathcal{I}^\phi_{EC}=\int d^{4}x\sqrt{g}[-\frac{1}{2}\nabla_\mu\phi\nabla^\mu\phi-\frac{1}{2}m^2\phi^2]\\
&\overset{\star} {\nabla} {}_\mu
\nabla^\mu\phi-m^2\phi=0,
\end{aligned}
\right.
\end{equation}
where $\overset{\star} {\nabla} {}_\mu=\nabla_\mu-K_\mu
$, with $K_\mu$ the trace of contortion tensor. It shows that the equation of motion in (\ref{b20}) is not the minimal coupling extension of $\partial_\mu\partial^\mu\phi-m^2\phi=0$. On the other hand, the situation stands even worse because it is not clear that which Lagrangian would correspond to equation $\nabla{}_\mu
\nabla^\mu\phi-m^2\phi=0$; or to say it radically, starting from the minimal coupled equation of motion for $\phi$, one is not able to deduce an analytically expressed Lagrangian. \par
This is not a problem only faced by scalar field. For vector field $A^\mu$(to avoid the gauge problem,we use massive vector field through out the whole paper) and Dirac field $\psi$, one finds
\begin{equation} \label{a0} 
\left\{
\begin{aligned}
&\mathcal{I}^A_{EC}=\int d^{4}x\sqrt{g}[-\frac{1}{4}F_{\mu\nu} F^{\mu\nu}-\frac{1}{2}m^2A_\mu A^\mu]\\
&\overset{\star} {\nabla} {}_\mu
F^{\mu\nu}-m^2A^\nu=0,
\end{aligned}
\right.
\end{equation}
where $F^{\mu\nu}={\nabla}^\mu A^\nu-{\nabla}^\nu A^\mu$, and
\begin{equation} \label{a2} 
\left\{
\begin{aligned}
&\mathcal{I}^\psi_{EC}=\int d^{4}x\sqrt{g}[\frac{i}{2}(\Bar{\psi}\gamma^\mu\nabla_\mu \psi-\nabla_\mu \bar{\psi}\gamma^\mu\psi)-m\bar{\psi}\psi]\\
&\frac{i}{2}\gamma^\mu{\nabla} {}_\mu
\psi+\frac{i}{2}\gamma^\mu\overset{\star} {\nabla} {}_\mu
\psi-m\psi=0\\
&\frac{i}{2}{\nabla} {}_\mu
\bar{\psi}\gamma^\mu+\frac{i}{2}\overset{\star} {\nabla} {}_\mu
\bar{\psi}\gamma^\mu+m\bar{\psi}=0.
\end{aligned}
\right.
\end{equation}
For both vector field and Dirac field, neither can we unify all the covariant derivatives in (\ref{a0}) and (\ref{a2}), nor finding a clearly expressed Lagrangian for the minimal-coupled equations of motion.
\par
It is notable that when treating Dirac field, it is tempting to construct the modified action

\begin{equation} \label{reviesd} 
\overcirc{\mathcal{I}}^\psi=\int d^{4}x\sqrt{g}[\frac{i}{2}(\Bar{\psi}\gamma^\mu\nabla_\mu \psi-\nabla_\mu \bar{\psi}\gamma^\mu\psi)-m\bar{\psi}\psi+\frac{i}{2}\bar{\psi}K_\mu \gamma^{\mu}\psi]
\end{equation}
which leads to
\begin{equation}
i\gamma^\mu{\nabla} {}_\mu
\psi-m\psi=0.
\end{equation}
This is the satisfactory minmal-coupled equation for $\psi$. But one can promptly calculate its conjugate equation from $\overcirc{\mathcal{I}}^\psi$:
\begin{equation}
i\overset{\star} {\nabla} {}_\mu
\bar{\psi}\gamma^\mu+m\bar{\psi}=0,
\end{equation}
which is not minimally coupled. Why there is a break of symmetric form between the conjugate equations? The point is that the standard action $\mathcal{I}^\psi_{EC}$ is a real function, while the quantity $\frac{i}{2}\bar{\psi}K_\mu \gamma^{\mu}\psi$ is a pure imaginary function, so  $\overcirc{\mathcal{I}}^\psi$ is a complex function. Such Lagrangian is illegal because it would add redundant constrains to the field. The direct consequence is that one can no longer deduce the equation of $\bar{\psi}$ form equation of $\psi$. The conjugate equation of $i\gamma^\mu{\nabla} {}_\mu
\psi-m\psi=0$ should be $i{\nabla} {}_\mu
\bar{\psi}\gamma^\mu+m\bar{\psi}=0$, but not $i\overset{\star} {\nabla} {}_\mu
\bar{\psi}\gamma^\mu+m\bar{\psi}=0$.

\par So the ambiguity arises in EC that we cannot make both the matter Lagrangian and the equation of motion minimally coupled. The difference in equation of motion would cause physical consequence thus one may seriously ask where shall the MCP start from, or is there a way to make the theory unique?\par
The key leads to this MCP ambiguity is that torsion plays a role in the covariant derivative: when proceeding the minimal action principle with a Lagrangian in EC, the quantity $\int d^4x\sqrt{g}\nabla{}_\mu B^\mu=\int d^4x\sqrt{g}(\widetilde{\nabla}_\mu+K_\mu) B^\mu$ is not a surface integral and does not vanish, where $B^\mu$ is an arbitrary vector field.\par
So far as the authors know, the MCP problem was first observed by Kibble\cite{Kibble:1961ba} and one can find a legible description by Saa\cite{Saa:1996mt}. To solve the problem Saa suggested modifying the volume element as $e^\theta\sqrt{g}d^4x$, where $\theta$ is an introduced scalar field and satisfies $\partial_\mu\theta=K_\mu$. In this model the quantity $\int d^4x e^\theta\sqrt{g}\nabla{}_\mu B^\mu$ turns out to be a surface term.
Another approach provided by 
Ka\'zmierczak\cite{Kazmierczak:2009jz,Kazmierczak:2009aq} was to write down a new spacetime connection in place of the original one： $\widehat{\Gamma}^\rho_{\mu\nu}=\Gamma^\rho_{\mu\nu}-\delta^\rho_\mu K_\nu$.This connection guarantees that  $\int d^4x\sqrt{g}\widehat{\nabla}{}_\mu B^\mu$ is a surface integral. In this paper we propose another possibility of modifying the space-time connection to solve the MCP problem, which we would present in section II. In section III, we come back to the models of Saa and Ka\'zmierczak, explain how they work, compare them with our model, and make some further discussion.

\section{Modified Coupling of gravitation field and matter fields}

We propose a new covariant derivative $\mathcal{D}$ which slightly alters the standard MCP:

\begin{equation} \label{3} 
\mathcal{D}_\mu B^\nu=\partial_\mu B^\nu+\mathbb{C}^\nu_{\lambda\mu}B^\lambda
\end{equation}
\begin{equation}
\mathcal{D}_\mu B_\nu=\partial_\mu B_\nu-\mathbb{C}^\lambda_{\nu\mu}B_\lambda,
\end{equation}

where $\mathbb{C}^\nu_{\lambda\mu}$ is the modified connection
\begin{equation} \label{4} 
\mathbb{C}^\nu_{\lambda\mu}=\Gamma^\nu_{\lambda\mu}-\frac{1}{3}(\delta^\nu_\mu K_\lambda-g_{\lambda\mu} K^\nu).
\end{equation}

With some algebra one can find:
\begin{equation} \label{5} 
\int d^4x\sqrt{g}\mathcal{D}_\mu B^\mu=d^4x\sqrt{g}[\widetilde{\nabla}_\mu +K_\mu-\frac{1}{3}(\delta_\nu^\nu K_\mu-g_{\mu\nu}K^\nu)]B^\mu
=\int d^4x\sqrt{g}\widetilde{\nabla}_\mu B^\mu=\int d^4x \partial_\mu(\sqrt{g} B^\mu).
\end{equation}\par
Such definition of  $\mathcal{D}_\mu$ fits all properties as a covariant derivative:\par
(a)Linearity
\begin{align} 
\nonumber
&\mathcal{D}_\lambda(\alpha E^\mu{}_\nu+\beta B^\mu{}_\nu)\\ \nonumber
=&
\partial_\lambda(\alpha E^\mu{}_\nu+\beta B^\mu{}_\nu)+\mathbb{C}^\mu_{\rho\lambda}(\alpha E^\rho{}_\nu+\beta B^\rho{}_\nu)-\mathbb{C}^\rho_{\nu\lambda}(\alpha E^\mu{}_\rho+\beta B^\mu{}_\rho)\\\nonumber
=&\alpha(\partial_\lambda E^\mu{}_\nu+\mathbb{C}^\mu_{\rho\lambda} E^\rho{}_\nu-\mathbb{C}^\rho_{\nu\lambda} E^\mu{}_\rho)+\beta(\partial_\lambda B^\mu{}_\nu+\mathbb{C}^\mu_{\rho\lambda} B^\rho{}_\nu-\mathbb{C}^\rho_{\nu\lambda} B^\mu{}_\rho)\\=&
\alpha\mathcal{D}_\lambda E^\mu{}_\nu+\beta\mathcal{D}_\lambda B^\mu{}_\nu
\end{align}
\par
(b)Leibniz rule
\begin{align} 
\nonumber&\mathcal{D}_\rho ( E^\mu{}_\nu B^\lambda )\\\nonumber=&\partial_\rho( E^\mu{}_\nu B^\lambda )+\mathbb{C}^\mu_{\epsilon\rho}( E^\epsilon{}_\nu B^\lambda )-\mathbb{C}^\epsilon_{\nu\rho}( E^\mu{}_\epsilon B^\lambda )+\mathbb{C}^\lambda_{\epsilon\rho}( E^\mu{}_\nu B^\epsilon )\\\nonumber=&
\partial_\rho E^\mu{}_\nu B^\lambda + \mathbb{C}^\mu_{\epsilon\rho} E^\epsilon{}_\nu B^\lambda -\mathbb{C}^\epsilon_{\nu\rho} E^\mu{}_\epsilon B^\lambda + E^\mu{}_\nu \partial_\rho B^\lambda+E^\mu{}_\nu\mathbb{C}^\lambda_{\epsilon\rho}  B^\epsilon\\=&\mathcal{D}_\rho  E^\mu{}_\nu B^\lambda + E^\mu{}_\nu\mathcal{D}_\rho  B^\lambda 
\end{align}
\par
(c)Contraction rule
\begin{align} 
\nonumber&\mathcal{D}_\rho T^{\mu\lambda}{}_\lambda\\\nonumber=&\partial_\rho T^{\mu\lambda}{}_\lambda+ \mathbb{C}^\mu_{\epsilon\rho} T^{\epsilon\lambda}{}_\lambda+\mathbb{C}^\lambda_{\epsilon\rho} T^{\mu\epsilon}{}_\lambda-\mathbb{C}^\epsilon_{\lambda\rho}T^{\mu\lambda}{}_\epsilon\\=&\partial_\rho T^{\mu\lambda}{}_\lambda+ \mathbb{C}^\mu_{\epsilon\rho} T^{\epsilon\lambda}{}_\lambda.
\end{align}
\par 
With the new covariant derivative in hand we can now build up the theory for scalar field, massive vector field and Dirac field respectively.\par

For scalar field, we have $\mathcal{D}_\mu\phi=\partial_\mu\phi$. The Lagrangian and the equation of motion reads:
\begin{equation}
\left\{
\begin{aligned}
&\mathcal{I}^\phi_{\mathbb{C}}=\int d^{4}x\sqrt{g}[-\frac{1}{2}\mathcal{D}_\mu\phi\mathcal{D}^\mu\phi-\frac{1}{2}m^2\phi^2]\\
&\mathcal{D}_\mu\mathcal{D}^\mu\phi-m^2\phi=0.
\end{aligned}
\right.
\end{equation}
\par
For massive vector field:

\begin{equation}
\left\{
\begin{aligned}
&\mathcal{I}^A_{\mathbb{C}}=\int d^{4}x\sqrt{g}[-\frac{1}{4}\mathcal{F}_{\mu\nu}\mathcal{F}^{\mu\nu}-\frac{1}{2}m^2A_\mu A^\mu]\\
&\mathcal{D}_\mu\mathcal{F}^{\mu\nu}-m^2A^\nu=0,
\end{aligned}
\right.
\end{equation}

where
\begin{equation} 
\mathcal{F}_{\mu\nu}=\mathcal{D}_\mu A_\nu-\mathcal{D}_\nu A_\mu.
\end{equation}
\par
  For Dirac field, 
 \begin{equation}
\left\{
\begin{aligned}
  &\mathcal{I}^\psi_{\mathbb{C}}=\int d^{4}x\sqrt{g}[\frac{i}{2}(\Bar{\psi}\gamma^\mu\mathcal{D}_\mu \psi-\mathcal{D}_\mu \bar{\psi}\gamma^\mu\psi)-m\bar{\psi}\psi]\\&i\gamma^\mu\mathcal{D}_\mu\psi-m\psi=0\\&i\mathcal{D}_\mu\bar{\psi}\gamma^\mu+m\bar{\psi}=0.
\end{aligned}
\right.
\end{equation}
  \\Expression of covariant derivative of a spinor field is given in appendix 1, and the calculation of variation is given in appendix 2.
\par
Each of the equation of motion above coincides with the corresponding one that extended form SR by replacing $\partial$ with $\mathcal{D}$. The structure of $\mathcal{D}$ removes the MCP ambiguity though it turns out to be a non-minimal coupling in Riemann-Cartan spacetime.

\section{discussion}
\subsection{Three treatments on the uniqueness problem}

 Let us now discuss a little bit about Saa's model. The modified volume element 
$e^\theta\sqrt{g}d^4x$ makes the integral  $\int d^4x e^\theta\sqrt{g}\nabla{}_\mu B^\mu$ a surface integral:

\begin{align}\nonumber
\int d^4x e^\theta\sqrt{g}\nabla{}_\mu B^\mu&=\int d^4x e^\theta\sqrt{g}[\widetilde{\nabla}_\mu B^\mu+K_\mu B^\mu]\\\nonumber&=\int d^4x \sqrt{g}[-(\widetilde{\nabla}_\mu e^\theta) B^\mu+K_\mu B^\mu e^\theta]\\\nonumber
&=\int d^4x \sqrt{g}[- e^\theta B^\mu\partial_\mu\theta+K_\mu B^\mu e^\theta]\\\nonumber
&=\int d^4x \sqrt{g}[- e^\theta B^\mu K_\mu +K_\mu B^\mu e^\theta]\\
&=0,
\end{align}
where one uses the condition that $e^\theta B^\mu$ vanish at infinite at the second line, and $\partial_\mu\theta=K_\mu$ at the fourth line.
 The model guarantees that a covariant divergence of a vector $\nabla_\mu B^\mu$ turns to be a surface term with the modified volume element, thus, one can freely integrate by parts with covariant derivative $\nabla$. But this model introduces an elementary field $\theta$ in order to produce contortion trace. The model also predicts that part of the torsion tensor propagates outside the matter distribution and shows inconsistencies to the known experiments\cite{Boyadjiev:1998uk,Fiziev:1998ga}.
\par
Compared with Saa's approach, Ka\'zmierczak does not introduce new field, and only slightly alters the spacetime connection: $\widehat{\Gamma}^\rho_{\mu\nu} \equiv \Gamma^\rho_{\mu\nu}-\delta^\rho_\mu K_\nu$. For this covariant derivative one has
\begin{align} \nonumber
\int d^4x \sqrt{g}\widehat{\nabla{}}_\mu B^\mu&=\int d^4x \sqrt{g}[\widetilde{\nabla{}}_\mu B^\mu+K_\mu B^\mu-\delta^\rho_\mu K_\rho B^\mu]\\\nonumber&=\int d^4x \sqrt{g}\widetilde{\nabla{}}_\mu B^\mu\\
&=\int d^4x \partial_\mu(\sqrt{g}B^\mu).
\end{align}
\\
By writing $\widehat{\nabla}_\mu B^\nu=\partial_\mu B^\nu+\Gamma^\nu_{\lambda\mu}B^\lambda-\delta^\nu_\lambda K_\mu B^\lambda=\partial_\mu B^\nu+\mathbb{N}^\nu_{\lambda\mu}B^\lambda$, one can prove that the derivative $\widehat{\nabla}$ retains the linearity, Leibniz rule and contraction rule just as in Section II. However, one can check that $\widehat{\Gamma}^\rho_{\mu\nu}$ is nolonger metric:\\\par
\begin{equation}
\widehat{\nabla}_\mu g_{\nu\lambda}=\widetilde{\nabla}_\mu g_{\nu\lambda}-K^\rho{}_{\nu\mu} g_{\rho\lambda}+\delta^\rho_\nu K_\mu g_{\rho\lambda}-K^\rho{}_{\lambda\mu} g_{\nu\rho}+\delta^\rho_\lambda K_\mu g_{\nu\rho}=2K_\mu g_{\nu\lambda}\not=0,
\end{equation}
\\where the properies $\widetilde{\nabla}_\mu g_{\nu\lambda}=0$(just as in GR) and $K_{\lambda\nu\mu}=-K_{\nu\lambda\mu}$ have been used.

\par In comparison, for our modified derivative $\mathcal{D}$, the connection $\mathbb{C}^\nu_{\lambda\mu}$  is still metric:
\begin{equation}
\mathcal{D}_\mu g_{\nu\lambda}=\partial_\mu g_{\nu\lambda}-\mathbb{C}^\rho_{\nu\mu}g_{\rho\lambda}-\mathbb{C}^\rho_{\lambda\mu}g_{\nu\rho}=\widetilde{\nabla}_\mu g_{\nu\lambda}-\mathbb{K}^\rho{}_{\nu\mu}g_{\rho\lambda}-\mathbb{K}^\rho{}_{\lambda\mu}g_{\nu\rho}=\widetilde{\nabla}_\mu g_{\nu\lambda}=0.
\end{equation}
The point is that the modified contortion tensor $\mathbb{K}_{\nu\lambda\mu}=K_{\nu\lambda\mu}-\frac{1}{3}(g_{\nu\mu} K_\lambda-g_{\lambda\mu} K_\nu)$ is anti-symmetric in the first two indices, and so keeping the covariant derivative of metric zero just as in EC.\par

\subsection{A new observation on EC, comparison of EC and our approach}

We now examine how different our theory is from EC. For scalar field:

\begin{align}\nonumber
\mathcal{I}^\phi_{\mathbb{C}}&=\int d^{4}x\sqrt{g}[-\frac{1}{2}g^{\mu\nu}\mathcal{D}_\mu\phi\mathcal{D}_\nu\phi-\frac{1}{2}m^2\phi^2]\\\nonumber
&=\int d^{4}x\sqrt{g}[-\frac{1}{2}g^{\mu\nu}\nabla_\mu\phi\nabla_\nu\phi-\frac{1}{2}m^2\phi^2]\\
&=\mathcal{I}^\phi_{EC}.
\end{align}
 Our model shows no differences form EC action, and so is the equation of motion. Further more, both $\mathcal{I}^\phi_{\mathbb{C}}$ and $\mathcal{I}^\phi_{EC}$ share the same expression with $\mathcal{I}^\phi_{GR}=\int d^{4}x\sqrt{g}[-\frac{1}{2}\widetilde{\nabla}_\mu\phi\widetilde{\nabla}^\mu\phi-\frac{1}{2}m^2\phi^2]$, thus they all lead to the same equation of motion: $\widetilde{\nabla}_\mu\widetilde{\nabla}^\mu\phi-m^2\phi=0$. This equation differs from  ${{\nabla}_\mu}{\nabla}^\mu\phi-m^2\phi=0$ which is built up with the complete covariant derivative in EC, by a term $K_\mu \nabla{}^\mu\phi$. This fact indicates that scalar field does not couple to torsion in the above mentioned theories, and the uniqueness problem would not arise in EC if we present the scalar theory with $\widetilde{\nabla}$(and also with $\mathcal{D}$). However, the  $\widetilde{\nabla}$ should be regarded as a modified derivative in EC just as $\mathcal{D}$, and more importantly, we can not unify vector and Dirac field with $\widetilde{\nabla}$ since it would simply reduce to GR.
 \par 
 
 For Dirac field, we notice a very interesting property that our modified action actually equals the EC action:
\begin{align} \nonumber
\mathcal{I}^\psi_{\mathbb{C}}&=\int d^{4}x\sqrt{g}[\frac{i}{2}(\Bar{\psi}\gamma^\mu\mathcal{D}_\mu \psi-\mathcal{D}_\mu \bar{\psi}\gamma^\mu\psi)-m\bar{\psi}\psi]\\\nonumber&=\int d^{4}x\sqrt{g}[\frac{i}{2}(\Bar{\psi}\gamma^\mu\nabla_\mu \psi-\nabla_\mu \bar{\psi}\gamma^\mu\psi)-m\bar{\psi}\psi-\frac{i}{24}\Bar{\psi}(e_{\mu a}K_b-e_{\mu b}K_a)\gamma^\mu\gamma^a\gamma^b\psi-\frac{i}{24}\Bar{\psi}(e_{\mu a}K_b-e_{\mu b}K_a)\gamma^a\gamma^b\gamma^\mu \psi]\\&=\mathcal{I}^\psi_{EC}.
\end{align}
This means that, as for the case of scalar field, one can re-arrange $\mathcal{I}^\psi_{EC}$ in a different form to avoid the ambiguity problem. 

Indeed, the property $\mathcal{I}^\psi_{\mathbb{C}}=\mathcal{I}^\psi_{EC}$ tells us both theories should yield the same equations of motion, as can be verified below:
\begin{align}\nonumber
&\frac{i}{2}\gamma^\mu{\nabla} {}_\mu
\psi+\frac{i}{2}\gamma^\mu\overset{\star} {\nabla} {}_\mu
\psi-m\psi
\\\nonumber=&\frac{i}{2}\gamma^\mu{\nabla} {}_\mu
\psi+\frac{i}{2}\gamma^\mu({\nabla} {}_\mu-K_\mu)
\psi-m\psi
\\\nonumber=&i\gamma^\mu{\nabla} {}_\mu
\psi-\frac{i}{2}\gamma^\mu K_\mu
\psi-m\psi
\\\nonumber=&i\gamma^\mu{\nabla} {}_\mu
\psi-\frac{i}{12}(4\gamma^b K_b+2\gamma^a K_a)-m\psi
\\\nonumber=&i\gamma^\mu[\partial_{\mu}\psi+\frac{1}{4}W_{ab\mu}\gamma^a\gamma^b\psi]-\frac{1}{4}\frac{i}{3}(e_{\mu a}K_b-e_{\mu b}K_a)\gamma^\mu\gamma^a\gamma^b\psi-m\psi
\\\nonumber=&i\gamma^\mu[\partial_{\mu}\psi+\frac{1}{4}(W_{ab\mu}-\frac{1}{3}(e_{\mu a}K_b-e_{\mu b}K_a))\gamma^a\gamma^b\psi]-m\psi
\\\nonumber=&i\gamma^\mu\mathcal{D}_\mu \psi-m\psi
\\=&0.
\end{align}

As a consistence check, the conjugate equation is:

\begin{align}\nonumber
&\frac{i}{2}{\nabla} {}_\mu
\bar{\psi}\gamma^\mu+\frac{i}{2}\overset{\star} {\nabla} {}_\mu\bar{\psi}\gamma^\mu
+m\bar{\psi}
\\\nonumber=&\frac{i}{2}{\nabla} {}_\mu
\bar{\psi}\gamma^\mu+\frac{i}{2}( {\nabla} {}_\mu-K_\mu)\bar{\psi}\gamma^\mu
+m\bar{\psi}\\\nonumber=&i{\nabla} {}_\mu
\bar{\psi}\gamma^\mu-\frac{i}{2}\bar{\psi}K_\mu \gamma^\mu+m\bar{\psi}\\\nonumber=&i{\nabla} {}_\mu
\bar{\psi}\gamma^\mu+\frac{i}{12}\bar{\psi}(-2K_b \gamma^b-4K_a\gamma^a)+m\bar{\psi}\\\nonumber=&i{\nabla} {}_\mu
\bar{\psi}\gamma^\mu+\frac{i}{12}\bar{\psi}(e_{\mu a}K_b -e_{\mu b}K_a)\gamma^a\gamma^b\gamma^\mu+m\bar{\psi}\\\nonumber=&i\mathcal{D}_\mu\bar{\psi}\gamma^\mu+m\bar{\psi}\\=&0.
\end{align}

\par For vector field, however, our theory shows concrete difference from EC:
\begin{align}\nonumber
\mathcal{I}^A_{\mathbb{C}}&=\int d^{4}x\sqrt{g}[-\frac{1}{4}\mathcal{F}_{\mu\nu}\mathcal{F}^{\mu\nu}-\frac{1}{2}m^2A_\mu A^\mu]\\\nonumber
&=\int d^{4}x\sqrt{g}[-\frac{1}{4}F_{\mu\nu}F^{\mu\nu}-\frac{1}{2}m^2A_\mu A^\mu-\frac{1}{6}(A_\mu K_\nu-A_\nu K_\mu)(F^{\mu\nu}+\frac{1}{3}A^\mu K^\nu)]\\&=\mathcal{I}^A_{EC}-\int d^{4}x\sqrt{g}[\frac{1}{6}(A_\mu K_\nu-A_\nu K_\mu)(F^{\mu\nu}+\frac{1}{3}A^\mu K^\nu)].
\end{align}

\par
By the above discussion we conclude that although the uniqueness problem can be solved partially within EC(what we meant by partially is that we are able to unify the covariant derivatives for scalar field and Dirac field, but not for vector field), it is somehow accidental and one still needs a satisfactory way to make the whole theory consistent. We proposed another possibility to solve the problem and the new coupling between matter and gravity differs from EC only in vector field. It is worth mentioning that the pure gravity part of action remains untouched compared to EC. As a further attempt, one may feel like to build up the curvature tensor through the commutator $[\mathcal{D}_\mu,\mathcal{D}_\nu]$ instead of $[\nabla_\mu,\nabla_\nu]$. The effort that we have made in this paper is aiming to give a hint on the guiding principle of how matter couples with gravitational field in a more general spacetime than Riemann, especially when spacetime torsion plays an important role.

\section*{Acknowledgments}
This work is supported by the China NSF via Grants No. 11535005 and No. 11275077.

\section*{Appendix 1: Definitions and Notations}
Covariant derivative of a vector in Riemann-Cartan spacetime:

\begin{equation} \label{a} 
\nabla_\mu B^\nu=\partial_\mu B^\nu+\Gamma^\nu_{\lambda\mu}B^\lambda,
\end{equation}

\begin{equation} \label{a} 
\nabla_\mu B_\nu=\partial_\mu B_\nu-\Gamma^\lambda_{\nu\mu}B_\lambda.
\end{equation}

Torsion tensor:
\begin{equation} \label{b} 
S^\mu{}_{\nu\rho}=\frac{1}{2}(\Gamma^\mu_{\nu\rho}-\Gamma^\mu_{\rho\nu}).
\end{equation}
Antisymmetry of torsion tensor:
\begin{equation} \label{d} 
S_{\mu\nu\lambda}=-S_{\mu\lambda\nu}.
\end{equation}

Split connection into Christoffel symbols and contortion tensor:

\begin{equation} \label{c} 
\Gamma^\nu_{\lambda\mu}=\widetilde{\Gamma}^\nu_{\lambda\mu}+K^\nu{}_{\lambda\mu}.
\end{equation}

Definition of contortion tensor:
\begin{equation} \label{d} 
K^\nu{}_{\lambda\mu}=S_{\lambda\mu}{}^\nu+S_{\mu\lambda}{}^\nu+S^\nu{}_{\lambda\mu}.
\end{equation}
Antisymmetry of contortion tensor:
\begin{equation} \label{d} 
K_{\mu\nu\lambda}=-K_{\nu\mu\lambda}.
\end{equation}

Torsion trace:
\begin{equation}  
S_\mu=S_{\nu\mu}{}^\nu.
\end{equation}

Contortion trace:
\begin{equation} \label{d} 
K_\mu=K_{\nu\mu}{}^\nu=2S_\mu.
\end{equation}

The new covariant derivative of Dirac field:

\begin{equation} \label{d} 
\mathcal{D}_\mu\psi=\partial_{\mu}\psi+\frac{1}{4}\mathbb{W}_{ab\mu}\gamma^a\gamma^b\psi,
\end{equation}
\begin{equation} \label{d} 
\mathcal{D}_\mu\bar{\psi}=\partial_{\mu}\bar{\psi}-\frac{1}{4}\bar{\psi}\mathbb{W}_{ab\mu}\gamma^a\gamma^b.
\end{equation}
Where $\mathbb{W}_{\nu\lambda\mu}=\widetilde{\omega}_{\nu\lambda\mu}+\mathbb{K}_{\nu\lambda\mu}$ is the modified spin connection, with $\widetilde{\omega}^\nu{}_{a\mu}$ the Levi-Civita spin connection, and $\mathbb{K}_{\nu\lambda\mu}=K_{\nu\lambda\mu}-\frac{1}{3}(g_{\nu\mu} K_\lambda-g_{\lambda\mu} K_\nu)$ the modified contortion tensor. $W_{\nu\lambda\mu}=\widetilde{\omega}_{\nu\lambda\mu}+K_{\nu\lambda\mu}$ is the spin connection in EC.
\par
The covariant derivative used in Ka\'zmierczak's work:
\begin{equation} \label{a} 
\widehat{\nabla}_\mu B^\nu=\partial_\mu B^\nu+\Gamma^\nu_{\lambda\mu}B^\lambda-\delta^\nu_\lambda K_\mu B^\lambda,
\end{equation}

\begin{equation} 
\widehat{\nabla}_\mu B_\nu=\partial_\mu B_\nu-\Gamma^\lambda_{\nu\mu}B_\lambda+\delta^\lambda_\nu K_\mu B_\lambda.
\end{equation}\par
Finally, the star derivative $\overset{\star} {\nabla} {}_\mu$ should be regarded as a symbol rather than a derivative, for it dose not satisfy Leibniz rule:

\begin{align} \nonumber
&\overset{\star} {\nabla} {}_\mu (E^\rho{}_\nu B^\chi)\\\nonumber=& {\nabla} {}_\mu(E^\rho{}_\nu B^\chi)-K_\mu E^\rho{}_\nu B^\chi\\\nonumber=&{\nabla} {}_\mu E^\rho{}_\nu B^\chi+ E^\rho{}_\nu {\nabla} {}_\mu  B^\chi-K_\mu E^\rho{}_\nu B^\chi\\  \not=&\overset{\star} {\nabla} {}_\mu E^\rho{}_\nu B^\chi+ E^\rho{}_\nu \overset{\star} {\nabla} {}_\mu B^\chi.
\end{align}

\section*{Appendix 2: Calculations of Equations of Motion}
For scalar field,vary the action with respect to $\phi$:

\begin{align}\label{98} \nonumber
\delta\mathcal{I}^\phi_{\mathbb{C}}&=\int d^{4}x\sqrt{g}[-\mathcal{D}_\mu\delta\phi\mathcal{D}^\mu\phi-m^2\phi\delta\phi]\\&=
\int d^{4}x\sqrt{g}[\mathcal{D}_\mu\mathcal{D}^\mu\phi-m^2\phi]\delta\phi.
\end{align}

For massive vector field, vary the action with respect to $A_\nu$:
\begin{align}\label{99} \nonumber
\delta\mathcal{I}^A_{\mathbb{C}}&=\int d^{4}x\sqrt{g}[-\frac{1}{2}\delta\mathcal{F}_{\mu\nu}\mathcal{F}^{\mu\nu}-m^2 A^\nu \delta A_\nu]\\\nonumber&=\int d^{4}x\sqrt{g}[-\frac{1}{2}(\mathcal{D}_\mu\delta A_\nu-\mathcal{D}_\nu\delta A_\mu)\mathcal{F}^{\mu\nu}-m^2 A^\nu\delta A_\nu]\\\nonumber&=\int d^{4}x\sqrt{g}[-\mathcal{F}^{\mu\nu}\mathcal{D}_\mu\delta A_\nu-m^2 A^\nu\delta A_\nu]\\&=\int d^{4}x\sqrt{g}[\mathcal{D}_\mu\mathcal{F}^{\mu\nu}-m^2 A^\nu]\delta A_\nu.
\end{align}

For Dirac field, vary the action with respect to $\bar{\psi}$:
\begin{align}\label{x}
\delta\mathcal{I}^\psi_{\mathbb{C}}=\int d^{4}x\sqrt{g}[\frac{i}{2}(\delta\Bar{\psi}\gamma^\mu\mathcal{D}_\mu \psi-\mathcal{D}_\mu \delta\bar{\psi}\gamma^\mu\psi)-m\delta\bar{\psi}\psi].
\end{align}
To calculate the variation, we have to prove that the quantity $\Delta=\int d^{4}x\sqrt{g}[\delta\Bar{\psi}\gamma^\mu\widetilde{\nabla}_\mu \psi+\widetilde{\nabla}_\mu \delta\bar{\psi}\gamma^\mu\psi]$ is a surface term:

\begin{align}\label{x}\nonumber
\Delta &=\int d^{4}x\sqrt{g}[\delta\Bar{\psi}\gamma^\mu\widetilde{\nabla}_\mu \psi+\widetilde{\nabla}_\mu \delta\bar{\psi}\gamma^\mu\psi]\\\nonumber&=\int d^{4}x\sqrt{g}e^\mu_a[\delta\Bar{\psi}\gamma^a\widetilde{\nabla}_\mu \psi+\widetilde{\nabla}_\mu \delta\bar{\psi}\gamma^a\psi]\\\nonumber&=\int d^{4}x\sqrt{g}e^\mu_a[(\partial_\mu\delta\Bar{\psi}-\frac{1}{4}\delta\Bar{\psi}\widetilde{\omega}_{cd\mu}\gamma^c\gamma^d)\gamma^a\psi +\delta\Bar{\psi}\gamma^a(\partial_\mu\psi+\frac{1}{4}\widetilde{\omega}_{cd\mu}\gamma^c\gamma^d\psi)]
\\\nonumber&=\int d^{4}x\sqrt{g}e^\mu_a[\partial_\mu(\delta\Bar{\psi}\gamma^a\psi)+\frac{1}{4}\delta\Bar{\psi}\widetilde{\omega}_{cd\mu}(\gamma^{a}\gamma^{c}\gamma^{d}-\gamma^{c}\gamma^{d}\gamma^{a})\psi]
\\\nonumber&=\int d^{4}x\sqrt{g}e^\mu_a[\partial_\mu(\delta\Bar{\psi}\gamma^a\psi)+\frac{1}{4}\delta\Bar{\psi}\widetilde{\omega}_{cd\mu}(2\eta^{ac}\gamma^{d}-2\gamma^{c}\eta^{ad})\psi]
\\\nonumber&=\int d^{4}x\sqrt{g}e^\mu_a[\partial_\mu(\delta\Bar{\psi}\gamma^a\psi)+\frac{1}{2}\delta\Bar{\psi}\widetilde{\omega}^{a}{}_{d\mu}\gamma^d\psi-\frac{1}{2}\delta\Bar{\psi}\widetilde{\omega}_{c}{}^{a}{}_{\mu}\gamma^c\psi]
\\\nonumber&=\int d^{4}x\sqrt{g}[e^\mu_a\partial_\mu(\delta\Bar{\psi}\gamma^a\psi)+\frac{1}{2}(e^\mu_a\widetilde{\omega}^{a}{}_{d\mu})\delta\Bar{\psi}\gamma^d\psi-\frac{1}{2}(e^{\mu a}\widetilde{\omega}^{d}{}_{a}{}_{\mu})\delta\Bar{\psi}\gamma_d\psi]
\\\nonumber&=\int d^{4}x\sqrt{g}[e^\mu_a\partial_\mu(\delta\Bar{\psi}\gamma^a\psi)+\frac{1}{2}e^\mu_a e^a_\rho (\partial_\mu e^\rho_d+\widetilde{\Gamma}^\rho_{\lambda\mu}e^\lambda_d)\delta\Bar{\psi}\gamma^d\psi-\frac{1}{2}e^{\mu a}e^d_\rho(\partial_\mu e^\rho_a+\widetilde{\Gamma}^\rho_{\lambda\mu}e^\lambda_a)\delta\Bar{\psi}\gamma_d\psi]
\\\nonumber&=\int d^{4}x\sqrt{g}[e^\mu_a\partial_\mu(\delta\Bar{\psi}\gamma^a\psi)+\frac{1}{2}(\partial_\rho e^\rho_d+g^{\mu\rho} \partial_\mu e_{\rho d}+\widetilde{\Gamma}^\rho_{\lambda\rho}e^\lambda_d-g^{\mu\lambda}\widetilde{\Gamma}^\rho_{\mu\lambda}e_{\rho d})\delta\Bar{\psi}\gamma^d\psi]
\\\nonumber&=\int d^{4}x\sqrt{g}\left\{e^\mu_a\partial_\mu(\delta\Bar{\psi}\gamma^a\psi)+\frac{1}{2}[\partial_\rho e^\rho_d+g^{\mu\rho} \partial_\mu e_{\rho d}+\widetilde{\Gamma}^\rho_{\lambda\rho}e^\lambda_d+(\widetilde{\Gamma}^\rho_{\lambda\rho} e^\lambda_d-\frac{1}{2}g^{\mu\lambda}e^{\sigma}_d\partial_\mu g_{\sigma\lambda}-\frac{1}{2}g^{\mu\lambda}e^{\sigma}_d\partial_\lambda g_{\sigma\mu})]\delta\Bar{\psi}\gamma^d\psi\right\}
\\\nonumber&=\int d^{4}x\sqrt{g}\left\{e^\mu_a\partial_\mu(\delta\Bar{\psi}\gamma^a\psi)+\frac{1}{2}[\partial_\rho e^\rho_d+g^{\mu\rho} \partial_\mu e_{\rho d}+\widetilde{\Gamma}^\rho_{\lambda\rho}e^\lambda_d+(\widetilde{\Gamma}^\rho_{\lambda\rho} e^\lambda_d-g^{\mu\rho}\partial_\mu e_{\rho d}+\partial_\rho e^\rho_d)]\delta\Bar{\psi}\gamma^d\psi\right\}
\\\nonumber&=\int d^{4}x\sqrt{g}\left\{e^\mu_a\partial_\mu(\delta\Bar{\psi}\gamma^a\psi)+(\partial_\rho e^\rho_d+\widetilde{\Gamma}^\rho_{\lambda\rho} e^\lambda_d)\delta\Bar{\psi}\gamma^d\psi\right\}
\\\nonumber&=\int d^{4}x\left\{\sqrt{g}e^\mu_a\partial_\mu(\delta\Bar{\psi}\gamma^a\psi)+\sqrt{g}\partial_\rho e^\rho_d\delta\Bar{\psi}\gamma^d\psi+\partial_\rho\sqrt{g} e^\rho_d\delta\Bar{\psi}\gamma^d\psi\right\}
\\&=\int d^{4}x\partial_\mu\left\{\sqrt{g}e^\mu_a(\delta\Bar{\psi}\gamma^a\psi)\right\}.
\end{align}
We can now compute the variation:
\begin{align}  \nonumber
\delta\mathcal{I}^\psi_{\mathbb{C}}&=\int d^{4}x\sqrt{g}[\frac{i}{2}(\delta\Bar{\psi}\gamma^\mu\mathcal{D}_\mu \psi-\mathcal{D}_\mu \delta\bar{\psi}\gamma^\mu\psi)-m\delta\bar{\psi}\psi]\\\nonumber&=\int d^{4}x\sqrt{g}\left\{\frac{i}{2}e^\mu_a(\delta\Bar{\psi}\gamma^a\partial_{\mu}\psi+\frac{1}{4}\delta\Bar{\psi}\gamma^a\mathbb{W}_{c d\mu}\gamma^c\gamma^d\psi-\partial_{\mu}\delta\bar{\psi}\gamma^a \psi+\frac{1}{4}\delta\bar{\psi}\mathbb{W}_{c d\mu}\gamma^c\gamma^d\gamma^a\psi)-m\delta\bar{\psi}\psi\right\}\\\nonumber&=\int d^{4}x\sqrt{g}\left\{\frac{i}{2}e^\mu_a[\delta\Bar{\psi}\gamma^a\partial_{\mu}\psi+\frac{1}{4}\delta\Bar{\psi}\gamma^a(\widetilde{\omega}_{cd\mu}+\mathbb{K}_{cd\mu})\gamma^c\gamma^d\psi-\partial_{\mu}\delta\bar{\psi}\gamma^a\psi+\frac{1}{4}\delta\bar{\psi}(\widetilde{\omega}_{cd\mu}+\mathbb{K}_{cd\mu})\gamma^c\gamma^d\gamma^a\psi]-m\delta\bar{\psi}\psi\right\}
\\\nonumber&=\int d^{4}x\sqrt{g}\left\{\frac{i}{2}e^\mu_a[\delta\Bar{\psi}\gamma^a\widetilde{\nabla}_{\mu}\psi+\frac{1}{4}\delta\Bar{\psi}\gamma^a\mathbb{K}_{cd\mu}\gamma^c\gamma^d\psi-\widetilde{\nabla}_{\mu}\delta\bar{\psi}\gamma^a\psi+\frac{1}{4}\delta\bar{\psi}\mathbb{K}_{cd\mu}\gamma^c\gamma^d\gamma^a\psi]-m\delta\bar{\psi}\psi\right\}
\\\nonumber&=\int d^{4}x\sqrt{g}\left\{\frac{i}{2}e^\mu_a[2\delta\Bar{\psi}\gamma^a\widetilde{\nabla}_{\mu}\psi+\frac{1}{4}\delta\Bar{\psi}\mathbb{K}_{cd\mu}(\gamma^a\gamma^c\gamma^d+\gamma^c\gamma^d\gamma^a)\psi]-m\delta\bar{\psi}\psi\right\}
\\\nonumber&=\int d^{4}x\sqrt{g}\left\{\frac{i}{2}e^\mu_a[2\delta\Bar{\psi}\gamma^a\widetilde{\nabla}_{\mu}\psi+\frac{1}{4}\delta\Bar{\psi}2\mathbb{K}_{cd\mu}(\gamma^a\gamma^c\gamma^d-\eta^{ac}\gamma^d+\eta^{ad}\gamma^c)\psi]-m\delta\bar{\psi}\psi\right\}
\\\nonumber&=\int d^{4}x\sqrt{g}\left\{\frac{i}{2}e^\mu_a[2\delta\Bar{\psi}\gamma^a\widetilde{\nabla}_{\mu}\psi+\frac{1}{2}\delta\Bar{\psi}\gamma^a\mathbb{K}_{cd\mu}\gamma^c\gamma^d\psi+\frac{1}{2}\delta\Bar{\psi}(\mathbb{K}{}_c{}^{a}{}_\mu-\mathbb{K}{}^a{}_{c\mu})\gamma^c\psi]-m\delta\bar{\psi}\psi\right\}
\\&=\int d^{4}x\sqrt{g}\left\{\delta\Bar{\psi}[i\gamma^\mu\mathcal{D}_\mu\psi-m\psi]\right\}.
\end{align}

%%%%%%%%%%%%%%%%%%%%%%%%%%%%%%%%%%%%%%%%%%%%%

\end{document}